\documentstyle[psfig,pra,aps]{revtex}

\newcommand{\ie}{\mbox{i.\,e.\,\ }}

\newcommand{\eg}{\mbox{e.\,g.\,\ }}
\newcommand{\egc}{\mbox{e.\,g.\,}}

\newcommand{\denop}{\widehat{\rho}}
\newcommand{\tr}{\textsf{Tr}}

\begin{document}
\draft
\title{Simple computer model for the quantum Zeno effect}
\author{David Wallace\thanks{Email address: david.wallace@merton.ox.ac.uk}}
\address{Centre for Quantum Computation, The Clarendon Laboratory, Oxford University, Parks Road, Oxford OX1 3PU, UK}
\date{\today}
\maketitle

\begin{abstract}
This paper presents a simple model for repeated measurement of a quantum system: the evolution of a free particle, simulated by discretising the particle's position.  This model is easily simulated by computer and provides a useful arena to investigate the effects of measurement upon dynamics, in particular the slowing of evolution due to measurement (the `quantum Zeno effect').  The results of this simulation are discussed for two rather different sorts of measurement process, both of which are (simplified forms of) measurements used in previous simulations of position measurement.  A number of interesting results due to measurement are found, and the investigation casts some light on previous disagreements about the presence or absence of the Zeno effect. 
\end{abstract}

\section{Introduction}

In quantum mechanics, observations of a system generally affect the state of that system.  When the system is evolving from one state to another, the effect of repeated observations may be to slow down the rate of evolution; this phenomenon was called the quantum Zeno effect by Misra and Sudarshan\cite{misrasudarshan}.

Misra and Sudarshan's original presentation of the quantum Zeno effect made use of the projection postulate, and has been criticised (see, \egc, \cite{fearnlamb,ballentine}) as unphysical on these grounds; more recent discussions (\egc, \cite{peres,schulman}) have tended to model the measurement device explicitly.  In general these models have demonstrated a slowing due to measurement: the measurement entangles a measuring device with the system, and so causes decay of off-diagonal elements in the density matrix of the system; this decay slows, but does not halt, the evolution.  A review of the Zeno effect is \cite{homewhitaker}, which also discusses its conceptual implications (which will not concern us here).

This paper is concerned with computer models of the Zeno effect, and in particular with measurement of the position of a one-dimensional particle.  This problem is interesting in that it simulates a \emph{continuous} measurement, and has been modelled by Fearn and Lamb\cite{fearnlamb} and by Altenm\"{u}ller and Schenzle\cite{as}; both treated a particle in a double-well potential and investigated the effects of measurement on the time taken for the particle to tunnel between the wells.  Interestingly, these two papers found opposite conclusions: Fearn and Lamb coupled the particle to a pointer, and found that this coupling actually increased the rate of penetration; Altenm\"{u}ller and Schenzle used a (simulated) laser to determine which well the particle was in (without measuring its position more precisely) and found a Zeno-type slowing of the tunnelling rate.

These computer simulations are relatively complex: Altenm\"{u}ller and Schenzle's model, in particular, involves very sophisticated simulation of the measurement interaction in addition to the need to solve the Schr\"{o}dinger equation numerically for the evolution of the particle.  The model presented here goes to the opposite extreme: it considers the simplest possible one-particle evolution, that of a free particle.  As such it is less `realistic' than the models mentioned above; on the other hand, it is conceptually simple enough to avoid being bogged down by details, while being complex enough to show a wide variety of interesting behaviour in response to measurement.  

The model itself is described in section \ref{model}, and the conventions for showing its output are described in section \ref{conv}. Sections \ref{sharp} and \ref{pointer} use the model to examine the effects of two very different types of measurement, qualitatively similar to Altenm\"{u}ller and Schenzle's coarse-grained measurement of position and to Fearn and Lamb's pointer measurement, respectively.  Section \ref{concl} draws some tentative conclusions.

\section{Description of the model}\label{model}

The model uses a periodic lattice to simulate a free particle.  The continuum of 
position states is replaced by N discrete states, labelled 
\[\left|X_0 \right\rangle,\left|X_1 \right\rangle,\left| X_2 \right\rangle,\ldots,
\left|X_{N\!-\!1}\right\rangle\]
and with the distance between successive states being $L$.  (For programming reasons 
N is chosen to be a power of 2; in the runs described here, 256 was used.)  
The analogues of the momentum states are then constructed by discrete Fourier 
transform:

\begin{equation}
\left|P_k\right\rangle =\frac{1}{\sqrt{N}}\sum_{n=0}^{N-1} 
\exp\left(-\frac{2 \pi i k n}
{N}\right) \left|X_n\right\rangle. 
\end{equation}
Those states with $k \leq N/2$ are intended to represent momentum states with 
momentum $2 \pi \hbar k/N L$ (note that $k$ is dimensionless).  The states with 
$k>N/2$ represent momentum states with momentum $- 2 \pi \hbar (N-k)/ N L$.

The free particle Hamiltonian is then

\begin{equation}
 \widehat{\textsf{H}} = (\hbar^2/\lambda L^2) \sum_{k=0}^{N-1} E(k) \left|P_k\right\rangle \left\langle P_k \right| 
\end{equation}
where

\[ E(k) = \left\{ \begin{array}{ll}  k^2/2 & (k \leq N/2) \\
 (N-k)^2 /2 & (k > N/2) \end{array} \right. \]
and $\lambda$ is a parameter analogous to inverse mass.  We are free to 
choose the length, time and mass units used to be such that $\hbar$, $L$ and 
$\lambda$ are numerically  equal to one, and this was done in the computer 
simulation.

Now we can solve the dynamics exactly: if $\left|\psi(0)\right\rangle = \left|P_k\right\rangle$ then

\begin{equation}
\left|\psi(t)\right\rangle = \exp(-iE(k)t) \left|P_k\right\rangle \label{mom_evolution}. 
\end{equation}
So given any density operator, we may find its evolution by decomposing it into momentum states and applying Eq.\,(\ref{mom_evolution}).  Formally,

\begin{equation}
\denop(0) = 
\sum_{k,k'=0}^{N-1} \left|P_{k}\right\rangle \left\langle P_{k}\right|\denop(0)\left|P_{k'}\right\rangle \left\langle P_{k'}\right|;
\end{equation}
hence

\begin{eqnarray}
\denop(t) & = & 
\sum_{k,k'=0}^{N-1} \exp\{it[E(k')-E(k)]\} \nonumber \\
& \times & \left|P_{k}\right\rangle \left\langle P_{k}\right|\denop(0)\left|P_{k'}\right\rangle \left\langle P_{k'}\right|
\label{rho_evolution}.
\end{eqnarray}
Measurements are specified in terms of the position basis, as is the initial 
state of the system (which was always taken to be pure).  To simulate measurement 
we suppose the system to have become entangled with some sort of measurement 
device; then we trace over the states of this device to get a reduced density 
matrix for the system.  (Note that this is the reverse of the usual
procedure, in which we would trace over the states of the system to find
the probability of various measurement outcomes.  Here, however, the
purpose of the measurement is not to find any information about the
system but simply to track the effects of measurement \emph{on} the
system.)
 
  If measurements are to be made after $T$ time units the process is as follows: \begin{enumerate}

\item Fourier-transform $\denop$ into the momentum basis

\item Time-evolve it by formula \ref{rho_evolution}, for a time $T$

\item Fourier-transform $\denop$ into the position basis

\item Carry out the measurement. The method, of course, varies according to the chosen measurement, but the effect on the particle will always be to modify the density matrix (in the position basis) in some relatively simple way (for instance, a perfect measurement of position would leave the diagonal elements untouched and set the off-diagonal elements to zero).

\end{enumerate}

Essentially all of the processing time was used carrying out the Fourier transforms, using the Fast Fourier Transform algorithm.

The reader may be wondering to what extent the periodic nature of momentum in the model affects its validity: after all, there is a sharp transition at $k=N/2$ between states intended to represent high rightwards momentum and those representing high leftwards momentum.  In one sense, of course, this does not matter as the model is still a perfectly legitimate quantum system in its own right; still, much of its interest is as a simulation of a particle with continuously varying position and artifacts such as Bragg reflection could spoil the validity of this simulation.
I do not believe this is a serious problem, for the following reason: we would expect the model's behaviour to deviate from that of a real particle only when the momentum distribution has spread significantly into the $k\sim N/2$ region.  If the spectrum remains concentrated away from the region --- as is the case throughout this paper, except for the eigenstate of `position' considered at the end of section \ref{sharp} --- we would not expect serious problems.
In any case, fears that artifacts are being generated as a result of the periodicity of momentum can usually be allayed by rerunning the simulation with an increased grid size.  For the measurements in this paper, this was done (increasing the grid from 256 to 512 points) for a number of test cases where the momentum distribution was significantly nonzero; no discernible differences occurred.

\section{Display conventions}\label{conv}

All of the diagrams below come in pairs.  The first diagram of each pair is a graph 
of the probability distribution of the system across the ``position'' eigenstates, 
i.\ e.\  the $x$-axis shows values of the dimensionless quantity
$n$ and the value of the graph at point $n$ is $\tr(\denop \left| X_n\right\rangle 
\left\langle X_n \right| )$.
  The second diagram is the probability distribution 
across the ``momentum'' eigenstates labelled by the dimensionless quantity $k$; 
since $k$ is only problematically 
proportional to momentum (see note at end of section~\ref{model}) the $x$-axis has 
itself been labelled ``k'' rather than ``momentum''.  The `time' variable has 
been multiplied by $10^3$ to produce manageable numbers.

It is important to remember that these graphs do not show the wavefunctions of 
pure states.  After a measurement, the state of the system is in general a highly 
incoherent density operator.

Although in the sequel these graphs will often be described as being probability 
distributions over position space, momentum space etc.\ it should be kept in mind 
that the system is finite-dimensional and periodic, and that the probability 
distribution is not really a function of a continuous variable. 

To save space, some of the results have been described verbally rather
than shown; the text indicates whenever a diagram has been omitted.

\section{Projection-valued measurements}\label{sharp}

To begin with we treat the simplest type of measurement: a sharp measurement of position (\ie a projection-valued measurement, or PVM).  We do not measure position exactly (clearly this is unphysical in the case of a real free particle) but coarse-grain the configuration space into $M$ regions and project the wave-function onto states with support in individual regions.  This is the original definition of measurement given by Von Neumann\cite{vonneumann}; in terms of the density operator, it amounts to setting to zero all elements $\denop_{lm}$ whenever $l$ and $m$ label states $\left|X_l\right\rangle$, $\left|X_m\right\rangle$ in different regions.  

However, we do not need to realise this `magically' by means of the projection postulate: it can be carried out physically by coupling the system to some measurement device which enters one of $M$ orthogonal states if the system is localised in one of the $M$ regions.  If the system is not localised in any of the regions, this process will entangle the system with the device, and as a result when we trace over the $M$ orthogonal `detector' states of the measurement device to get the reduced density operator of the system, this density operator will be modified as above.  This sort of measurement is similar in spirit to (but much more simplistic than) the measurement modelled by Altenm\"{u}ller and Schenzle.

In the example shown here, position space was divided into 6 regions
of equal size (with a small leftover region).  The boundaries between regions are shown on the diagrams by vertical lines.

\begin{figure} 
\vspace{0.4in} 
\psfig{figure=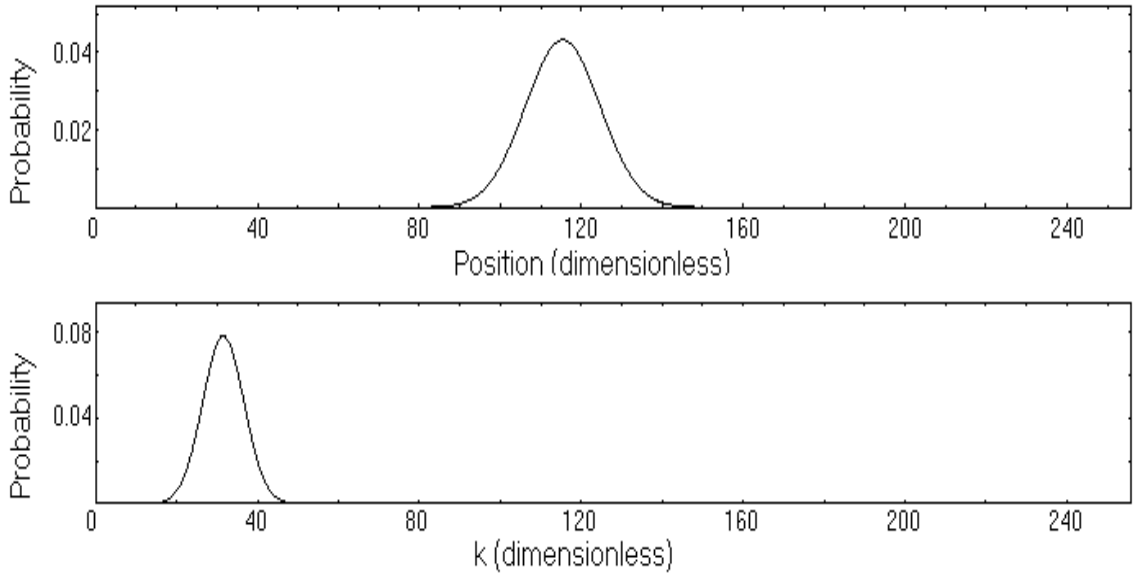,height=9.0cm,width=17.8cm,angle=0}  
\caption{Initially Gaussian state, width 8, momentum +31; time 400; no measurement} 
\label{fig1}
\end{figure}

The first choice of initial state was a Gaussian pure state located near the left\footnote{Of course, left here is arbitrary; the space is periodic.} of the position space and with a moderate rightward momentum:
\[\left|\psi\right\rangle = {\cal N} 
\sum_{n=1}^N 
\exp \left(\frac{(n-n_0)^2}{w^2} + \frac{2 \pi i k_0 n}{N}\right)\left|X_n\right\rangle\]
where $\mathcal{N}$ is a normalisation constant ($n_0$ was taken as 8; $k_0$ as 31). 
As expected, in the absence of measurement this packet moves rightwards and spreads out slightly; its state at time 400 is shown in figure~\ref{fig1}. The momentum distribution is of course unchanged.

\begin{figure}
\vspace{0.4in}
\psfig{figure=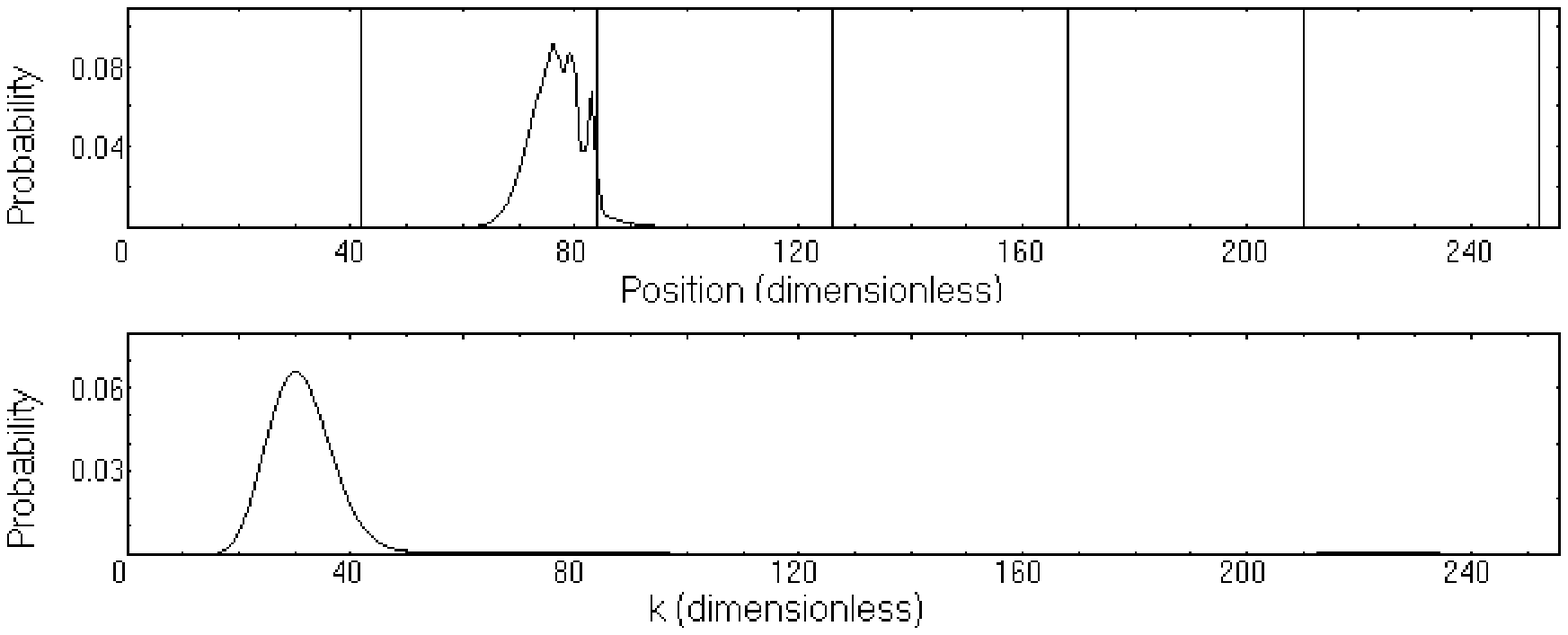,height=9.0cm,width=17.8cm,angle=0}
\caption{Initially Gaussian state, width 8, momentum +31; time 100; PVM once per unit time}
\label{fig2}
\end{figure}

The system was now rerun with a PVM being carried out at intervals of 1 unit.  
So long as the packet remains within a single measurement region its evolution 
proceeds normally.  At about time 60 (not shown) it reaches the boundary of the region, 
and at time 80 (not shown) small distortions have appeared at the rightmost edge of the packet. 
By time 100 these distortions have become magnified to large, jagged peaks (figure~\ref{fig2}), with a sharp drop in probability density to the right of the boundary: this suggests a Zeno-type effect is retarding the packet's motion through the boundary.

\begin{figure}
\vspace{0.4in}
\psfig{figure=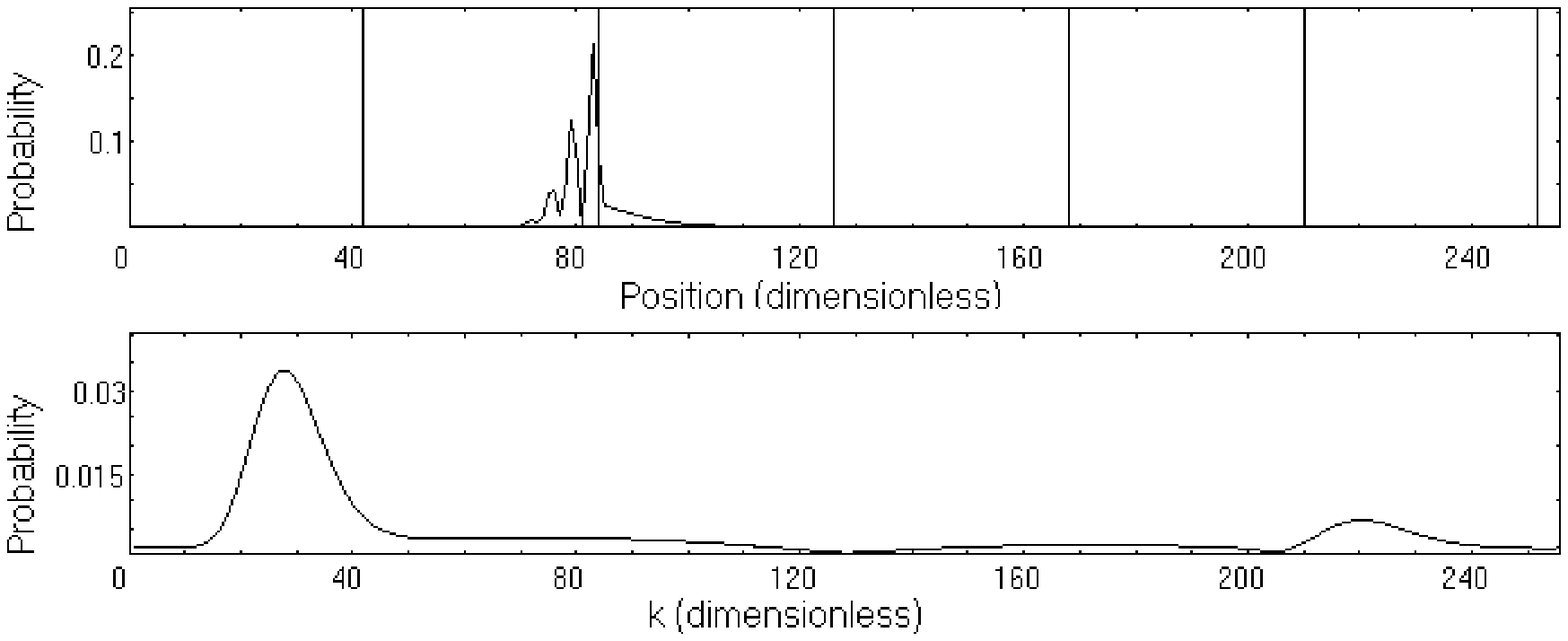,height=9.0cm,width=17.8cm,angle=0}
\caption{Initially Gaussian state, width 8, momentum +31; time 140; PVM once per unit time}
\label{fig3}
\end{figure}

\begin{figure}
\vspace{0.4in}
\psfig{figure=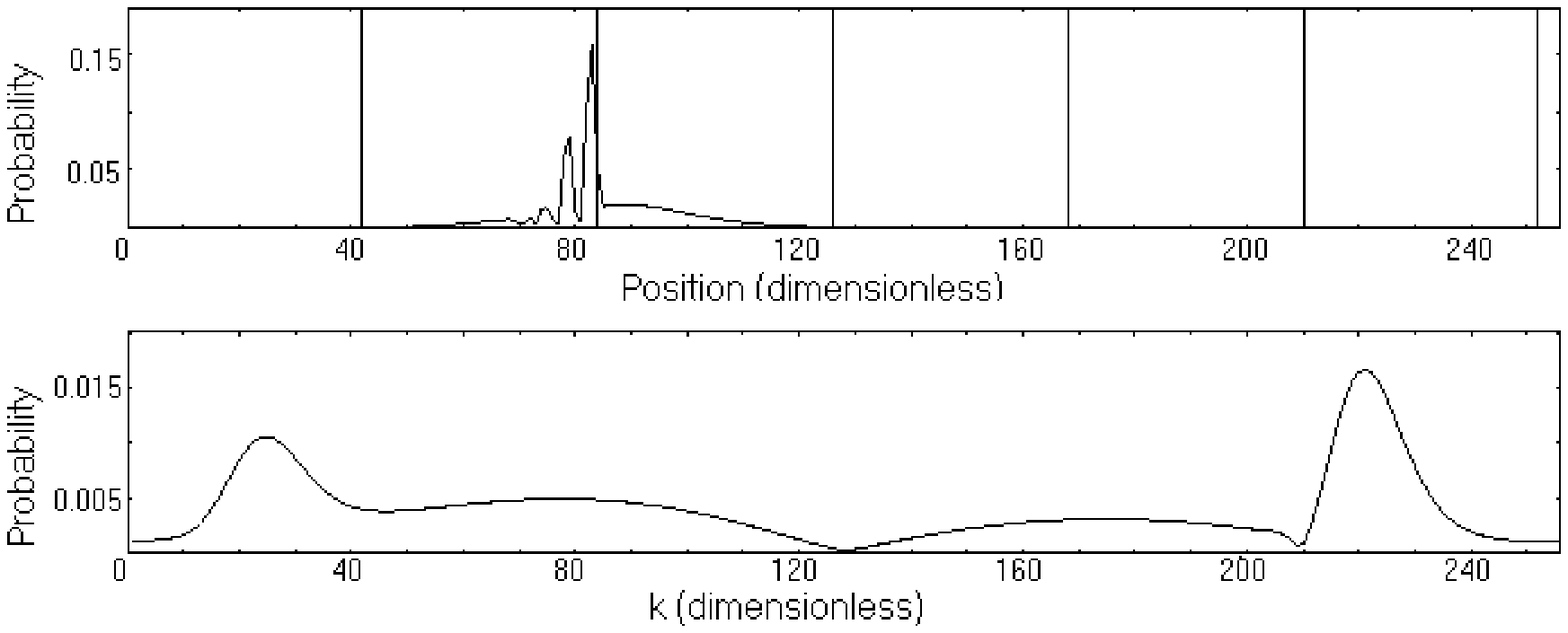,height=9.0cm,width=17.8cm,angle=0}
\caption{Initially Gaussian state, width 8, momentum +31; time 180; PVM once per unit time}
\label{fig4}
\end{figure}

By time 140 (figure~\ref{fig3}) the Gaussian shape has been entirely replaced by a series of jagged peaks to the left of the boundary, and a long, low tail to the right.  The momentum graph also shows that a small part of the packet has effectively been reflected, and now has negative momentum.  The fraction of the packet which has been reflected grows rapidly; by time 180 (figure~\ref{fig4}) it is significantly larger than the incident part of the packet.  Meanwhile, the tail to the right of the boundary has spread most of the way to the next boundary, remaining fairly smooth without the jagged structure to the left of the boundary.  It is also possible at this stage to see a small tail beginning to spread back to to the left, towards the other boundary of the initial region.

\begin{figure}
\vspace{0.4in}
\psfig{figure=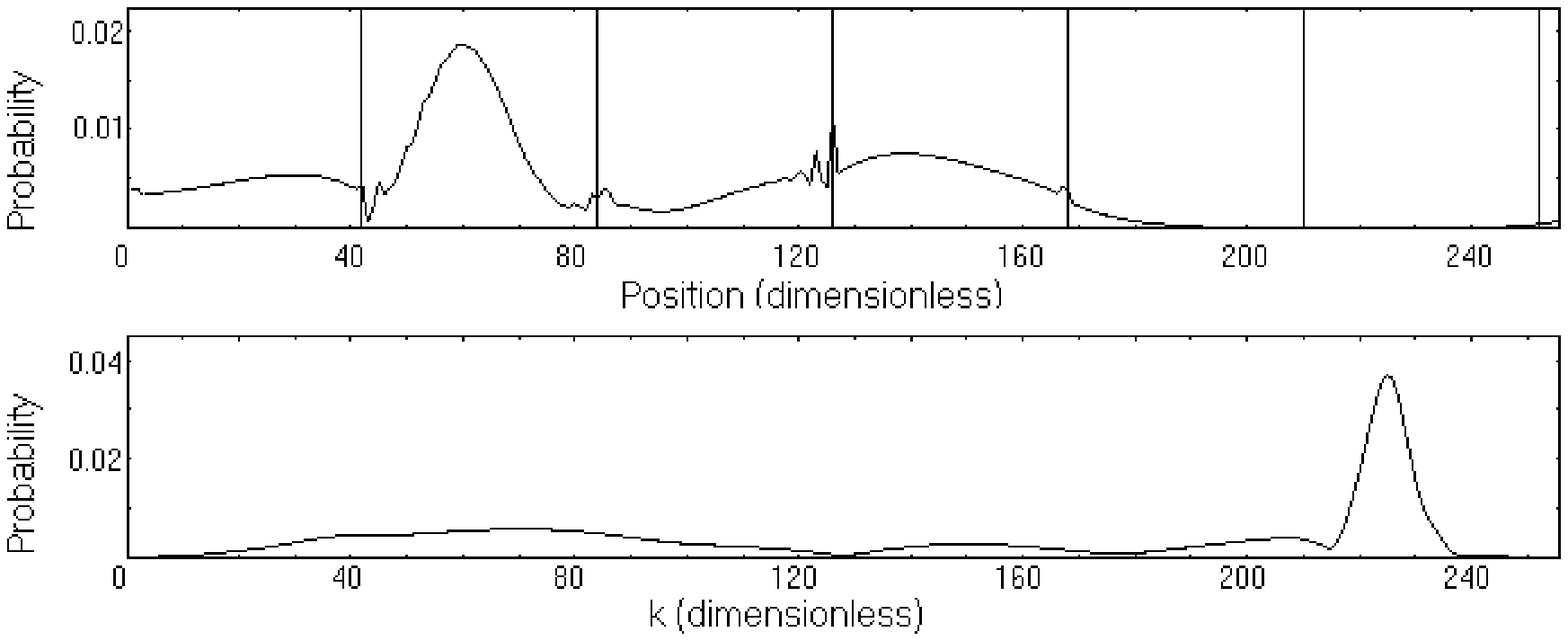,height=9.0cm,width=17.8cm,angle=0}
\caption{Initially Gaussian state, width 8, momentum +31; time 360; PVM once per unit time}
\label{fig5}
\end{figure}

By time 200 (not shown) the incident part of the momentum distribution has become 
very spread out.  By time 240 (not shown) the left and right tails of the distribution have grown significantly and are starting to interact with further measurement boundaries; the jagged part of the distribution shows signs of smoothing out and moving back off to the left.  By time 360 (figure~\ref{fig5}) this part has smoothed out completely into a leftwards-moving packet (presumably with the bulk of the negative-momentum part of the momentum spectrum) and is beginning to encounter the left measurement boundary.  The rest of the packet has continued to spread out to left and right, and shows some signs of interaction at other measurement boundaries.  Comparison of figure~\ref{fig5} with figure~\ref{fig1} shows that the state's rightward motion has been significantly slowed.

There is a striking resemblance to the a priori totally different problem of a wave-packet incident on a potential barrier.  It is very natural to interpret the diagrams as the position and momentum distributions of a pure state which collides with a barrier, is partially transmitted (the right-edge tail), mostly reflected (the negative-momentum peak), interferes with itself on reflection (the jagged region) and eventually is split up with the smaller part heading off to the right and the larger part, reformed into a rough wave-packet shape, moving to the left and eventually being incident on another barrier.  This resemblance is all the more surprising when we remember that the above is a completely false description: the graphs do not refer to a pure state at all, and the ``packet'' moving off to the left in figure~\ref{fig5} is highly incoherent.    

Variation of the number of measurement regions and the frequency of measurements produced no surprises: increasing either caused the Zeno freezing to increase.

\begin{figure}
\vspace{0.4in}
\psfig{figure=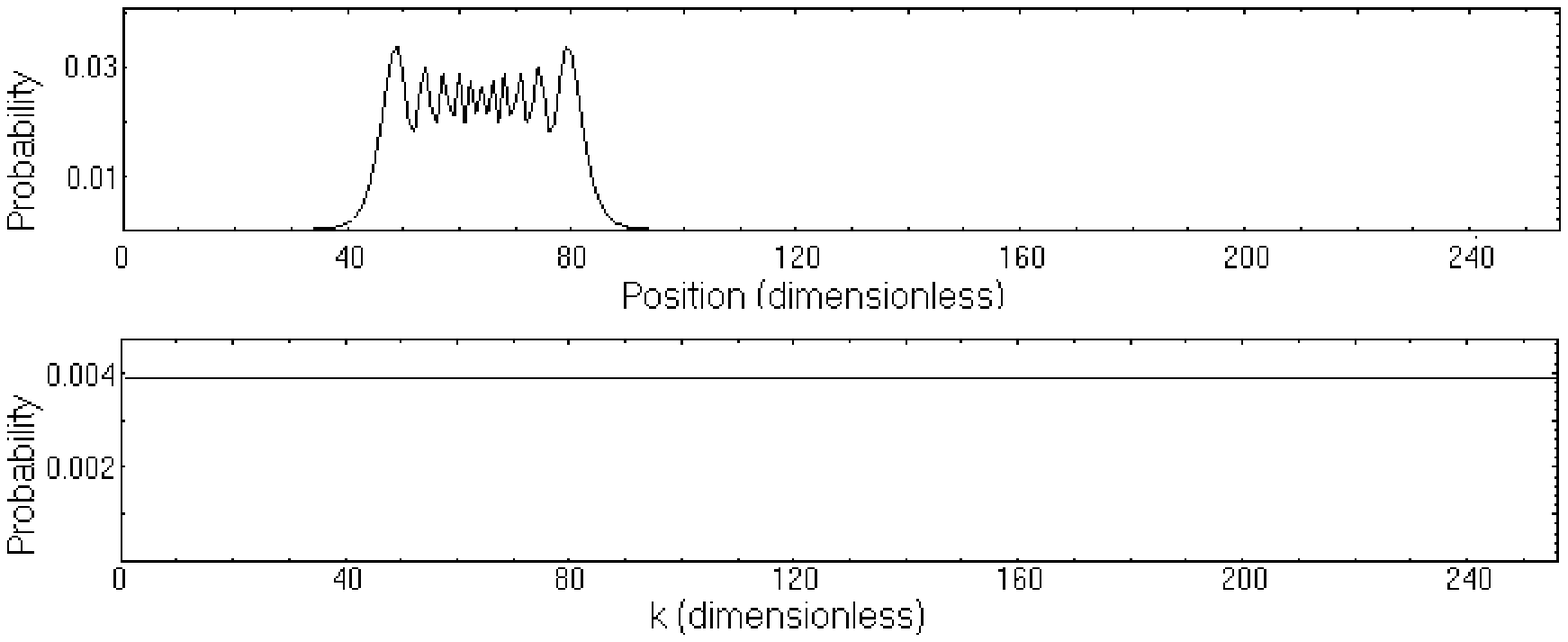,height=9.0cm,width=17.8cm,angle=0}
\caption{Initially position eigenstate; time 40; no measurement}
\label{fig6}
\end{figure}

\begin{figure}
\vspace{0.4in}
\psfig{figure=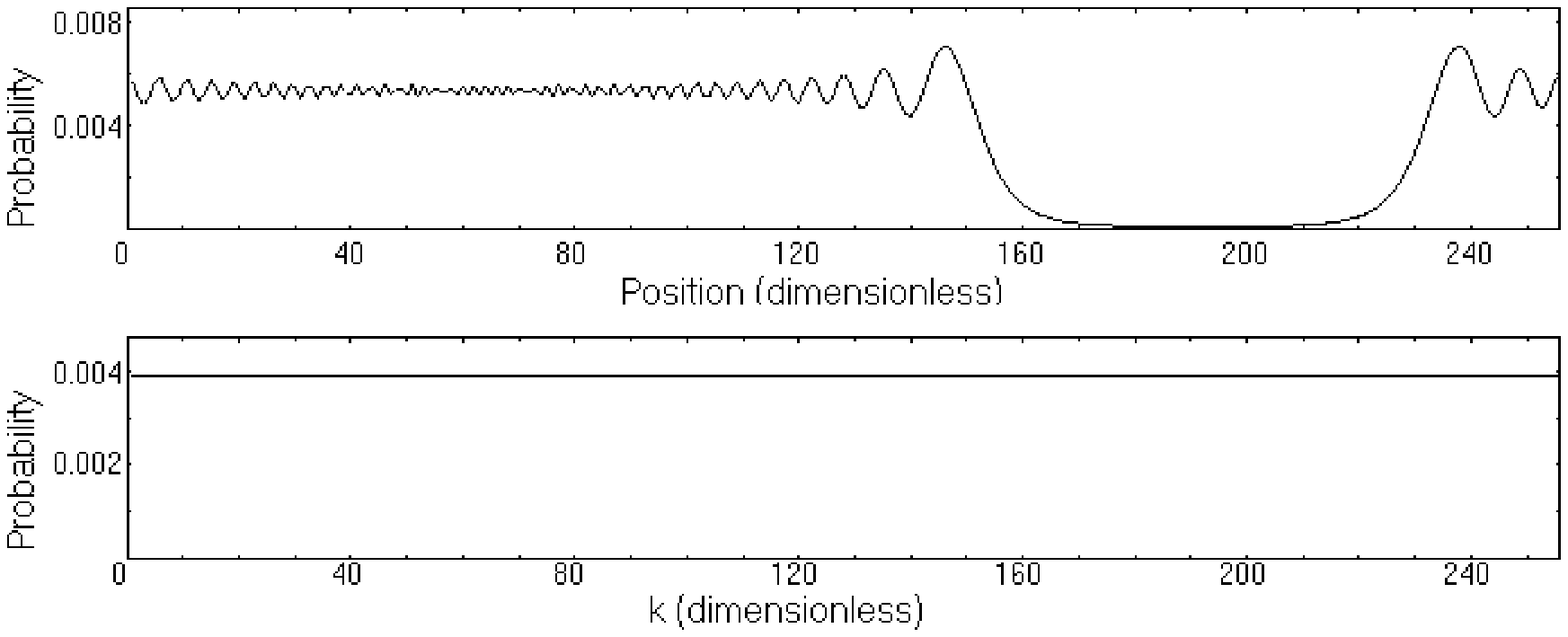,height=9.0cm,width=17.8cm,angle=0}
\caption{Initially position eigenstate; time 180; no measurement}
\label{fig7}
\end{figure}

\begin{figure}
\vspace{0.4in}
\psfig{figure=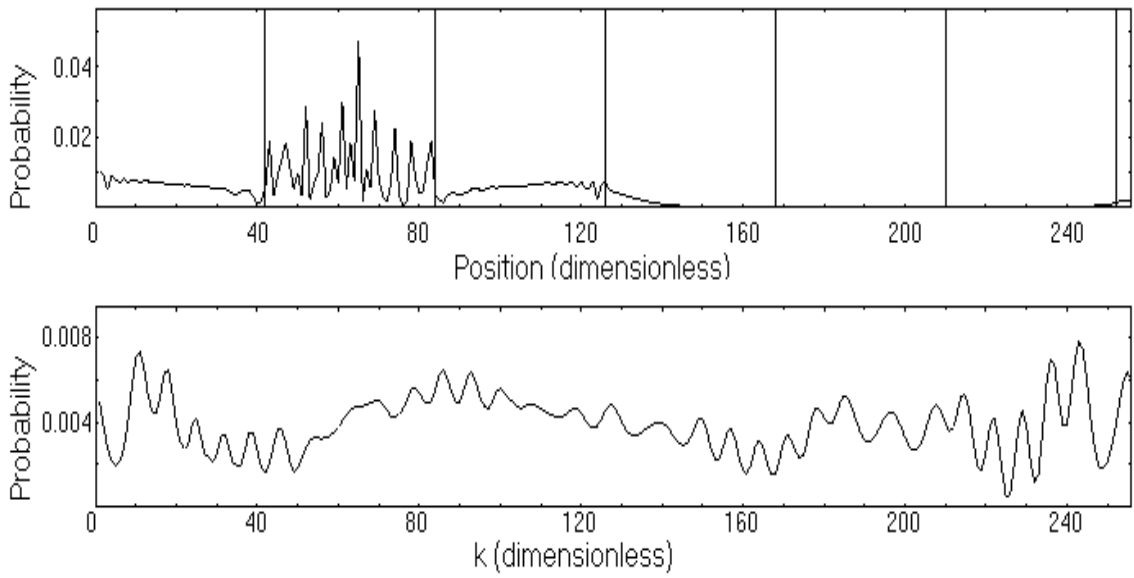,height=9.0cm,width=17.8cm,angle=0}
\caption{Initially position eigenstate; time 180; PVM once per unit time}
\label{fig8}
\end{figure}

As a second example --- and perhaps one more akin to decay --- the same  type of 
measurement was tried using an eigenstate of position as the initial state.  Such 
a state has no well-defined momentum but will tend to spread out fairly rapidly 
and to form sharp peaks which eventually soften (figures~\ref{fig6},\ref{fig7}).  
With a measurement, this process is sharply halted at the measurement boundary.  
The initial region continues to be filled with jagged peaks, but a relatively 
smooth tail leaks out of the region on either side (not shown).  This tail spreads out and in due course (figure~\ref{fig8}) interacts with the next set of measurement barriers; it behaves very similarly to the wave packets considered previously.  The bulk of the state remains confined within the initial region, and remains very unevenly distributed.  This resembles the behaviour at the measurement boundary for the wave-packet above.  Comparison of the measured system at time 180 (figure~\ref{fig8}) with the unmeasured system (figure~\ref{fig7}) shows that the system's evolution has been sharply reduced.
(It is important to note that this second example does not reliably simulate any continuous position measurement, as the momentum spectrum is completely delocalized and so there is no way to avoid the problems of Bragg reflection mentioned at the end of section \ref{model}.)

\section{Pointer measurements}\label{pointer}

In this section we consider an alternative model of measurement, in which a pointer is coupled to the particle's position.  This model was first constructed by Von Neumann\cite{vonneumann} and was developed further by Arthurs and Kelly\cite{arthurskelly} and by Lamb\cite{lamb} and Fearn and Lamb\cite{fearnlamb}; it is the measurement model used in their computer simulations.  See also \cite{peres} for a discussion. The measurement process is as follows: we take the pointer as being described by a one-dimensional wave-packet, initially taken to be $\chi(x)$, and take the measurement dynamic to be
\[\left|X_n\right\rangle\chi(x) \longrightarrow \left|X _n\right\rangle\chi(x-n),\]
so that, for a fairly localised pointer (\eg a sharply pointed Gaussian) if the system is in an eigenstate of position then the pointer will point out that eigenstate.  

If the pointer is sufficiently sharp as to be narrower than the gap between eigenstates, this process effectively reduces to a sharp measurement of position (\ie a PVM as in the last section, but with $M=N$).  However, the needle can only be this narrow by exploiting the discrete nature of position in the simulation.  To simulate a genuine measurement of position the pointer wave-function must be significantly wider than the gap between eigenstates.  In this case, the wave-functions of pointer states corresponding to closely spaced particle-position states would overlap, leading to imprecise measurements and a reduction of decoherence.

Here we consider the special case of a Gaussian pointer, with width $1/\alpha$ (\ie wave-function $\chi(x) = \chi_0 \exp(-\alpha^2 x^2/2)$).  The effect on the density matrix $\denop_{mn}$ of the particle of coupling it to this pointer is easily shown to be
\[\denop_{mn} \longrightarrow \denop_{mn} \exp(-\alpha^2 (m-n)^2/4).\]

As noted by Fearn and Lamb, measurements of this nature tend to cause the momentum distribution of the state to spread out (they claim that this momentum spreading increases the chance of barrier penetration in their model, and so gives the reverse of a Zeno effect).  However, it is also easy to show (see appendix) that the \emph{expected} momentum of the state is conserved by this sort of measurement interaction.  As such, moving wave-packets are not ideal choices of initial state, since their evolution is dominated by bulk motion and the measurement cannot really affect this motion.  This was in fact found when such states were put into the model: on measurement by a pointer, the packet spreads out but its centre-of-mass motion is not affected at all\footnote{In fact some slight slowing was eventually observed but this is an artefact of the model: when the momentum is sufficiently spread out the periodic nature of momentum space causes the high-positive-momentum states also to be high-negative-momentum states.  In a periodic system like this one momentum can be defined, and is conserved, only in a pragmatic sense.}.

\begin{figure}
\vspace{0.4in}
\psfig{figure=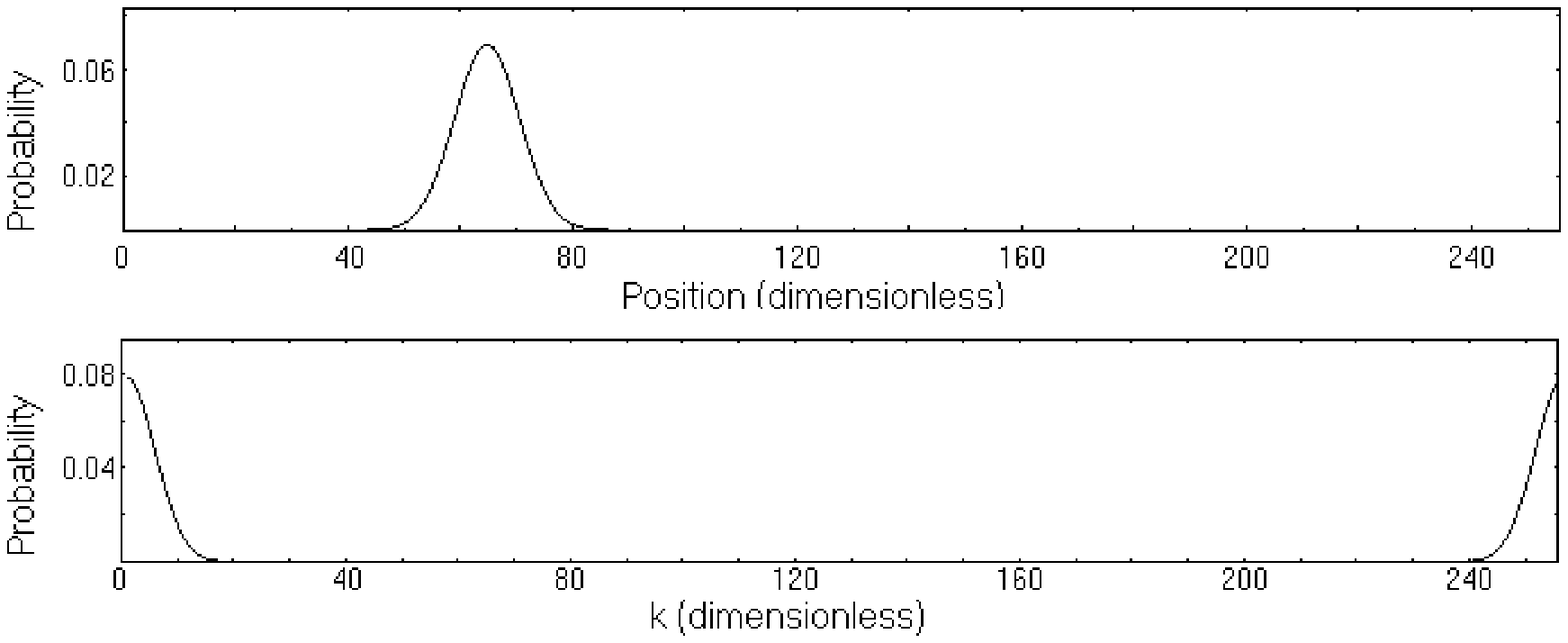,height=9.0cm,width=17.8cm,angle=0}
\caption{Initially Gaussian state, width 8, momentum 0; time 200; no measurement}
\label{fig9}
\end{figure}

To avoid this problem, a stationary Gaussian wavepacket was used as an initial state.  Such a state will evolve entirely by spread of the wave-packet; as this is a significantly slower process than bulk motion the timescale of measurement was reduced accordingly. Figure~\ref{fig9} shows the undisturbed evolution of this state at time 200.

\begin{figure}
\vspace{0.4in}
\psfig{figure=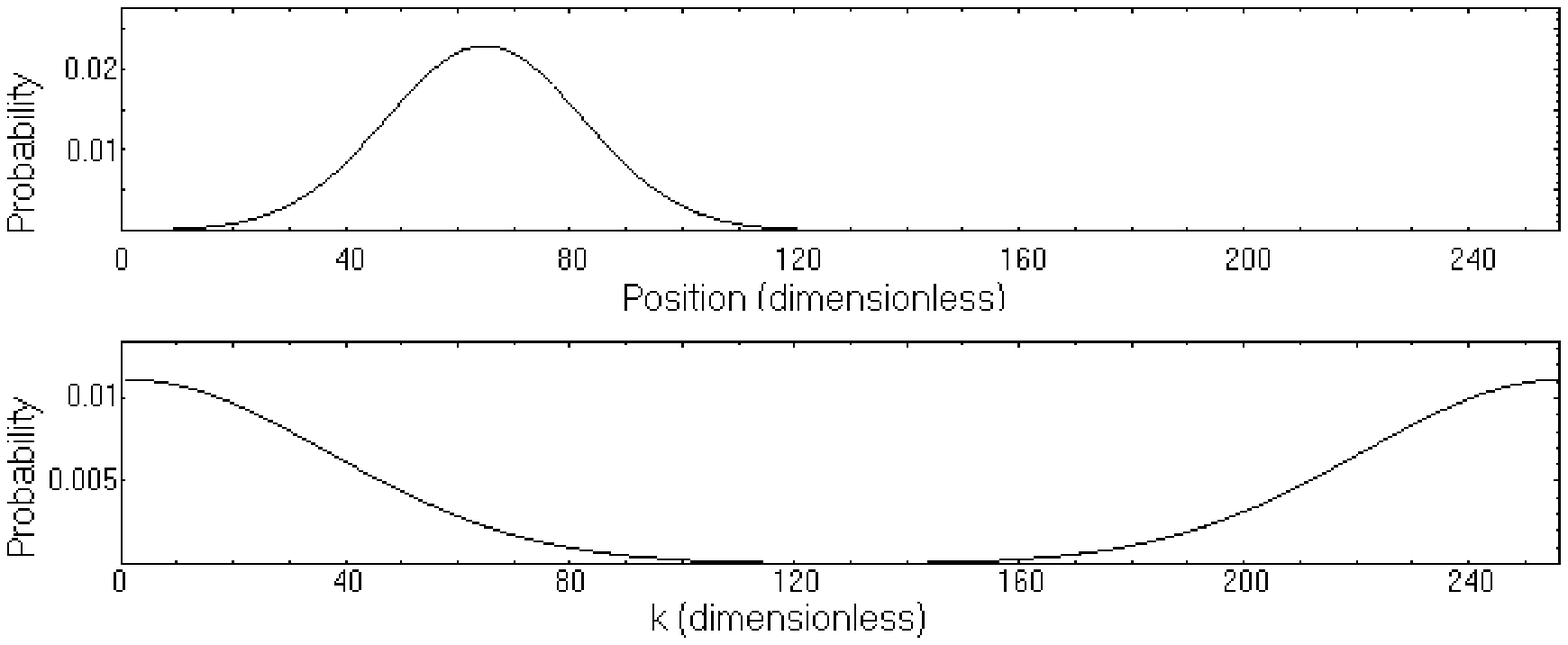,height=9.0cm,width=17.8cm,angle=0}
\caption{Initially Gaussian state, width 8, momentum 0; time 200; pointer measurement, pointer width 5, once per 10 unit time}
\label{fig10}
\end{figure}

\begin{figure}
\vspace{0.4in}
\psfig{figure=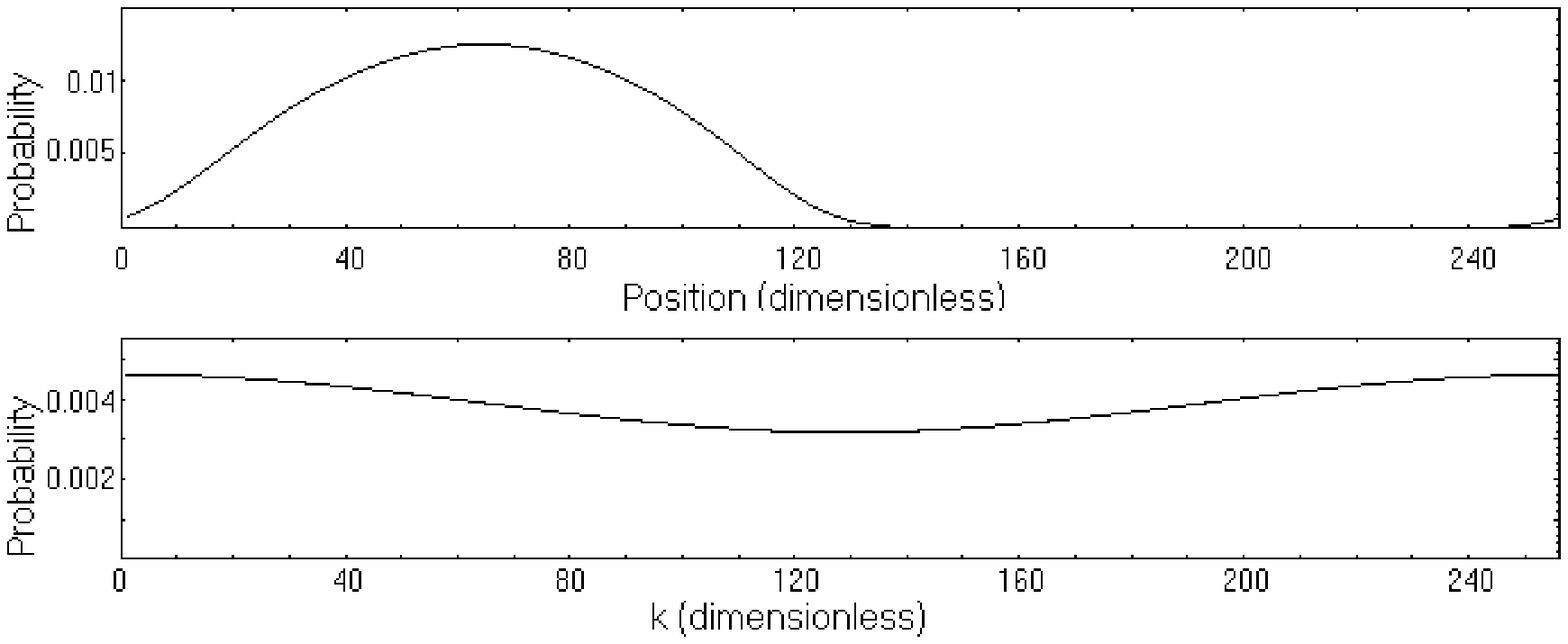,height=9.0cm,width=17.8cm,angle=0}
\caption{Initially Gaussian state, width 8, momentum 0; time 200; pointer measurement, pointer width 2, once per 10 unit time}
\label{fig11}
\end{figure}

A Gaussian needle of half-width 5 was then used to measure the system at intervals of 10 units.  The effect is significant (figure~\ref{fig10}): the momentum distribution is rapidly spread out and the position distribution soon follows suit, leading to a much faster evolution with measurement.  Increasing the sharpness of the pointer simply intensifies this effect (figure~\ref{fig11}).

This seems in accordance with Fearn and Lamb's results: a pointer measurement causes momentum spread and so, far from slowing the evolution of a system, it accelerates it.  

\section{Conclusions}\label{concl}

The model presented here is a quantum system with several hundred degrees of freedom, but whose evolution --- with or without measurement --- is straightforwardly calculated and understood, and which can be measured in a wide variety of ways.  As such it seems to be a useful half-way house between highly complex, detailed simulations on the one hand and simple, analytical models on the other.

Whether it is a \emph{realistic} quantum system is another matter: it is intended to model a free particle but clearly cannot do so perfectly.  However, provided that we are careful
\begin{enumerate}
\item to take the number of discrete positions large enough (see last paragraph of section \ref{model});
\item not to make measurements which distinguish adjacent positions; and
\item to avoid situations where the periodic nature of the position and momentum become relevant
\end{enumerate}
there seems no reason not to treat it as realistic. 

What can we learn from the model about the Zeno effect?  The main conclusion I would draw is that repeated measurement of a system has dynamical effects that are far more complex than just slowing the evolution.
Some sorts of measurements seem to produce slowing, while others actually accelerate the evolution; also novel effects (like the potential-barrier-like behaviour of the measurement in section \ref{sharp}) are observed, and 
changes of initial state (such as changing from a moving to a stationary packet in section \ref{pointer}) can make the difference between significant measurement-caused effects and few or no effects.

Fearn and Lamb, and Altenm\"{u}ller and Schenzle, found measurement to have opposite effects to one another in their respective computer models, and it is easy to read this as a genuine disagreement about whether the quantum Zeno effect is real.  It is satisfying to observe qualitatively the same effects in our model, giving support to a more conciliatory view: quantum measurement involves entangling a measurement device with the system to be measured, and this is bound to affect the dynamics of the system; but there are many ways to measure position, and it is perfectly plausible that different sorts of measurement will have different effects.

Investigating the details of how measurements affect the dynamics of systems lies rather beyond the scope of this paper, whose purpose is simply to show the complex sorts of phenomena which can occur and to present a simple model with which to investigate them.  The phenomena are clearly due to measurement-induced decoherence of the initially pure state; an obvious next step in investigating them would be to look at how rapidly the state does decohere (the separate plots of position and momentum given here are not ideal for this purpose; some sort of joint plot, such as the Wigner function(\cite{wigner}; discussed in \cite{peres}) would be more suited).  Nonetheless there seem to be interesting effects to investigate here, which may be obscured by the details of more complex models and lost in the over-simplifications of analytic ones.

\section*{Acknowledgements}

I am very grateful to Ian Aitchison for extensive discussions and constructive criticism.

\appendix

\section*{Proof that pointer measurements conserve momentum}

We work in the continuum limit. Recall that we are coupling the system to a Gaussian needle of half-width $1/\alpha$, so that the density matrix evolves like
\[\denop(X,X') \longrightarrow \denop(X,X') \exp(-\alpha^2 (X-X')^2/4).\]
Then the expected momentum is
\begin{eqnarray*}
<\widehat{\textsf{P}}> =  \tr{\widehat{\textsf{P}}\denop} & = &\int dX\ \left\langle X \right| \widehat{\textsf{P}}\denop \left| X \right\rangle 
\nonumber \\
& = & \int \mathrm{d}X\, \mathrm{d}X'\, \left\langle X \right| \widehat{\textsf{P}} \left| X' \right\rangle  \left\langle X' \right| \denop \left| X \right\rangle .
\end{eqnarray*}
Now
\[
\left\langle X \right| \widehat{\textsf{P}}\left| \psi \right\rangle = - i \frac{d}{dX} \langle X | \psi \rangle
\]
so
\[
\left\langle X \right| \widehat{\textsf{P}} \left| X' \right\rangle =-i \delta'(X-X')
\]
and therefore
\begin{eqnarray*}
<\widehat{\textsf{P}}>' & = & -i \int \mathrm{d}X\, \mathrm{d}X'\, \delta'(X-X') \denop'(X',X)
\\
& = & -i \int \mathrm{d}X\, \left. \frac{d\denop'}{dX'} \right|_{(X,X)}.
\end{eqnarray*}
We have
\begin{eqnarray*}
\frac{d\denop'}{dX'} & = & \frac{d}{dX'}
\left(\denop e^{-\alpha^2(X-X')^2/4}\right)
\\
& = &
\frac{d\denop}{dx'} e^{-\alpha^2(X-X')^2/4}
\\
& & + \denop . \frac{-\alpha^2}{2}(X-X') e^{-\alpha^2(X-X')^2/4}
\end{eqnarray*}
so
\[
\left.\frac{d\denop'}{dX'}\right|_{(X,X)}
=
\left.\frac{d\denop}{dX'}\right|_{(X,X)}.
\]
Hence the expected momentum is unchanged by the measurement; in fact this result holds for any pointer (not just a Gaussian one) provided it is symmetric about the centre and is differentiable.  Any sensible pointer should satisfy these requirements.

\bibliography{zc}
\bibliographystyle{prsty}

\end{document}